\documentclass[twoside]{JINST}
\usepackage{graphicx}
\usepackage{rangecite}
\usepackage{amsmath}
\usepackage[figuresright]{rotating}

\title{Study of ionization signals in a TPC filled with a mixture of liquid Argon and Nitrogen}

 \author{A. Ereditato, M.~Hess, S.~Jano\^{s}, I.~Kreslo, L.~Martinez, M.~Messina, U.~Moser, B.~Rossi, H.-U.~Sh\"{u}tz, M.~Zeller \\
 \llap{}Laboratory for High Energy Physics, University of Bern, Bern, Switzerland \\
}

\abstract{In this paper we report on the evidence for ionization track signals from cosmic ray muons and Compton electrons in a Time Projection Chamber (TPC) filled with liquid Argon and doped with different fractions of Nitrogen. This study has been conducted in view of the possible use of liquid Argon/Nitrogen TPCs for the detection of gamma rays in the resonant band of the Nitrogen absorbtion spectrum, a promising technology for security and medical applications.}

\begin{document}

\section{Introduction}
Time Projection Chambers (TPC) based on liquid noble gases as ionization and target medium are being developed since 1977 \cite{Rubbia77}. Among the main advantages of such detectors there are high-density, high filling factor, high-granularity and high spatial resolution. TPCs filled with liquid Argon are promising because of the availability and the low cost of Argon, and the proven high performance (see for example \cite{Ikarus1}). Sub-milimeter space resolution can be achived with such detectors for minimum ionizing particles, with full 3D reconstruction of the events.

A promising new technology proposed for the vast area of security applications is the Gamma Nuclear Resonant Absorbtion (GNRA) radiography \cite{GRA1,GRA2}. It assumes the use of high-resolution tracking detectors with a substantial content of Nitrogen as gamma-to-charged particle conversion medium. An appealing approach that we are proposing is to use liquid Argon TPCs for this purpose, doped with liquid Nitrogen to a concentration that provides high gamma conversion efficiency in the resonant band on one side, and sufficient charge transport and collection performance on the other. We plan to continue the related research and development of such a detector in near future by exploiting the outstanding imaging and particle identification capabilities of liquid Argon TPCs.

The first observations of signals in ionization chambers filled with mixtures of liquid Argon-Nitrogen date from 1948 \cite{DavidsonLarsh}. A notable effect of the Nitrogen concentration on the ionization pulse height was reported in that work. However, in 1963 it was shown that the charge attachment cross section in liquid Argon-Nitrogen mixtures is rather low, so that the efficient transport of the ionization signal in such mixtures is possible \cite{Swan}. Important studies on the production and transport of ionisation signals in liquid Argon in presence of Nitrogen and Oxygen impurities were reported in 1976 \cite{Hofmann}. The studied concentrations were up to 1750~ppm and 60~ppm for Nitrogen and Oxygen respectively. The strong effect due to Oxygen was observed while the effect of Nitrogen was much lower. An effect on the scintillation light yield in liquid Argon in presence of small amounts of Nitrogen was also mentioned in \cite{Berset}

In these works Nitrogen was considered as undesired contamination to the noble Argon, and therefore the concentrations studied were below $1\%$. In order to have sufficient gamma conversion efficiency the concentration of Nitrogen in the tracking medium must be of the order of 5-10$\%$ or more. The investigation of ionization charge transport in liquid Argon-Nitrogen mixtures with a substantial content of Nitrogen (from $7\%$ to $25\%$) was reported in 1984 \cite{Barabash}. It was shown that Nitrogen does not reduce the detected charge as strongly as Oxygen. However, the presence of Nitrogen can catalize the process of electron attachment on electronegative impurities, such as Oxygen, due to enhanced electron thermalisation. In \cite{Sakai} the possibility of the detection of ionization signals in mixtures with relatively high (up to $20 mol. \%$) concentrations of Nitrogen in liquid Argon was reported, together with measurements of the electron drift velocity. These results have shown the possibility to achieve reasonably efficient transport of the primary ionization charge packets in liquid Argon-Nitrogen mixtures over distances of the order of millimeters.
In this paper we study charge production and recombination in a prototype TPC filled with liquid Argon-Nitrogen mixtures with Nitrogen content of up to 6$\%$.

\section{Experimental setup}

\begin{figure}[htbp]	
\center\includegraphics[width=0.8\textwidth]{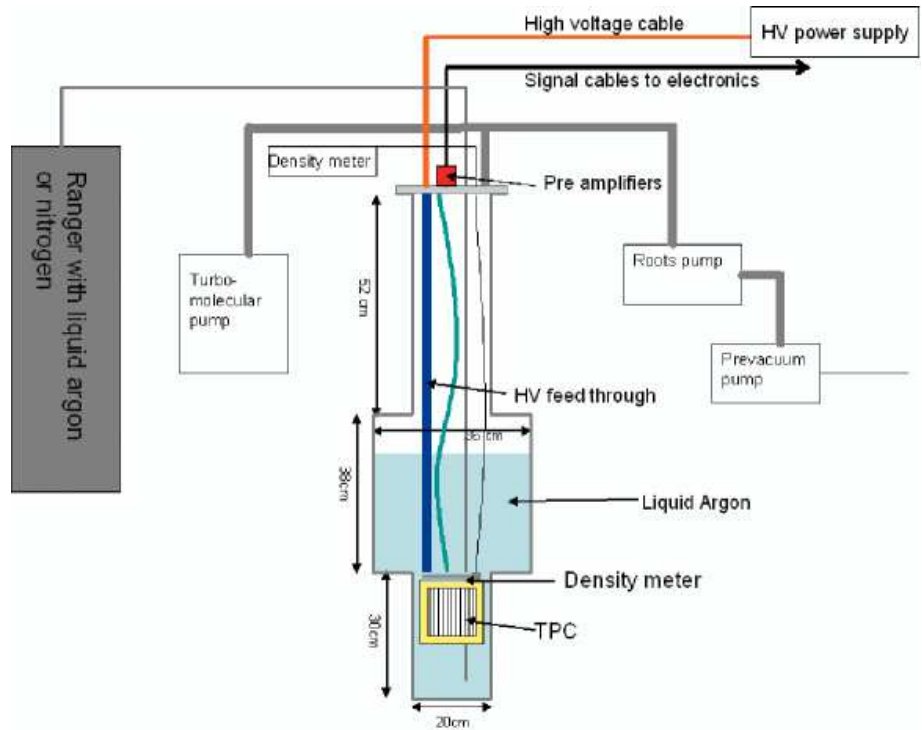}
\caption{Sketch of the experimental set-up.}\label{Fig:shematot}
\end{figure}

\begin{figure}[htbp]	
\center\includegraphics[width=0.60\textwidth]{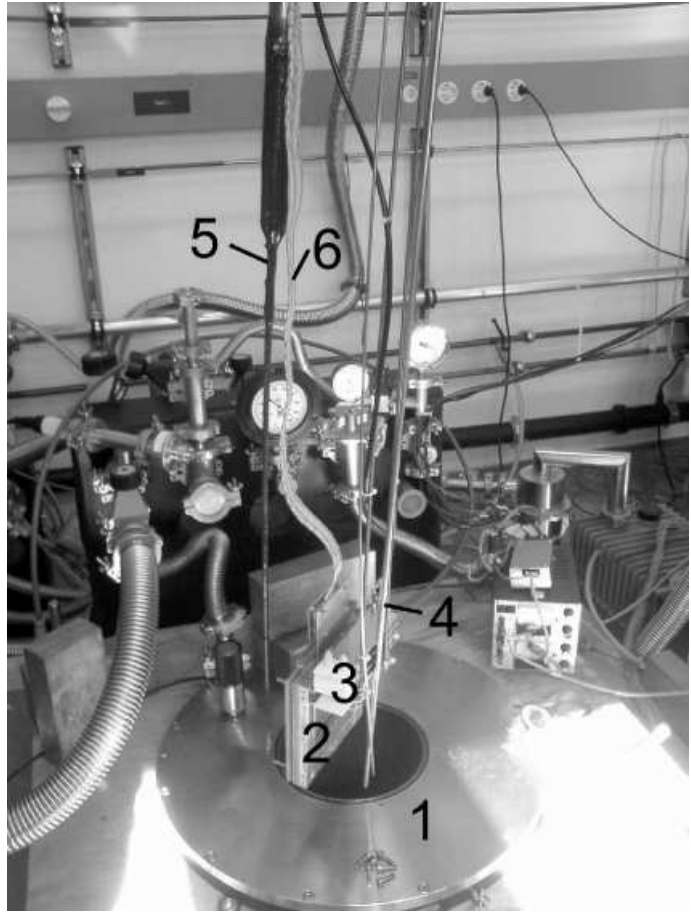}
\caption{Photograph of the whole experimental set-up (the TPC is out of the Dewar, the top flange is not visible): 1-Dewar, 2-TPC, 3-density sensor, 4-liquid Argon feed-in, 5-HV feedthrough, 6-signal cable.}\label{Photo1}
\end{figure}

A short drift-gap TPC was constructed for this study, with relatively high (about 10 kV/cm) electrostatic field. The chamber is placed in a hermetic vacuum-isolated volume (Dewar). High-voltage input and wire signals output are provided at 
the top flange of the Dewar with dedicated feedthrough connectors (Figures \ref{Fig:shematot}, \ref{Photo1}). 

The TPC is a wire chamber whith the cathode plate placed at a high negative potential and the wire plane (anode) at a potential near ground. The whole detector is completely immersed in liquid Argon. A charged particle passing through the liquid Argon volume ionizes the atoms close to the particle track.  Electrons are separated from the positive ions by the electric field and drift to the wires, while the ions drift to the cathode.  The electrons are then captured by the wires, hence producing an electric pulse. As a source of ionization, Compton electrons produced by $\gamma$ radiation from a $^{60}Co$ source are utilized as well as cosmic ray muons. 

\begin{figure}[htbp]	
\center\includegraphics[angle=90, width=0.9\textwidth]{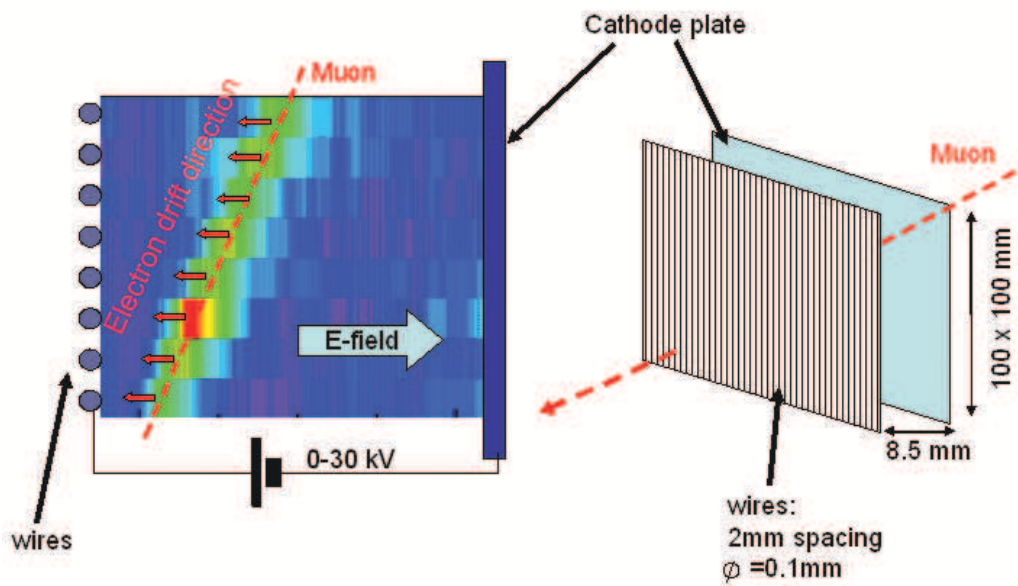}
\caption{TPC geometry.}\label{TPC}
\end{figure}

\begin{figure}[htbp]	
\center\includegraphics[width=0.7\textwidth]{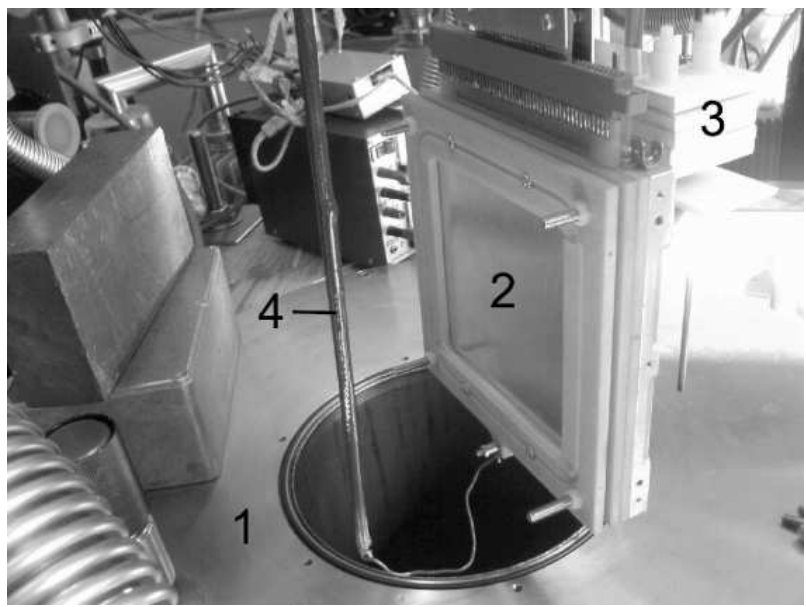}
\caption{The TPC and its Dewar: 1-Dewar, 2-TPC (cathode side), 3-density sensor, 4-HV feedthrough.}\label{Photo3}
\end{figure}

The chamber has a drift distance of 8.5 mm, 20 wires with 2 mm spacing and a length of 103 mm. Out of 20 wires 8 are read out and the others just connected to ground potential to provide a uniform field in the drift region. This results in an active volume of 14 $cm^3$ (Figures \ref{TPC} and \ref{Photo3}). 

In order to achieve the required level of liquid Argon purity the Dewar was first evacuated to a residual pressure of $1.4 \times 10^{-3}~mbar$. Then it was flushed with the pure cold gas Argon to remove remaining contaminations. The filling system was sealed and evacuated as well. The whole TPC setup was assembled with the use of low-degassing materials, such as aluminum, stainless steel, fiberglass. The high-voltage feedthrough and signal cables were isolated with PVC and polyurethane. However, their degassing at liquid Argon temperature may be neglected. The Argon used to fill the detector is a commercial liquified gas of a grade 48 ($99.998\%$). The Nitrogen used for doping has a grade 50 ($99.9990\%$). Both liquified gases are provided by CARBAGAS \footnote{CARBAGAS,Hofgut, 3073 G\"{u}mligen, Switzerland.}. The Argon-Nitrogen mixture is produced in the Dewar by adding measured quantities of liquid Nitrogen into the Dewar, filled with a known amount of liquid Argon. In order to monitor the Nitrogen concentration a simple sensor was developed, based on the measurement of the density of the liquified gas. 

The collected charge signals from the wires are amplified by low-noise integrating shaping amplifyers and read out by high-speed VME ADC modules. The readout chain for each wire consists of a charge-integrating pre-amplifier and a high-speed ADC channel. The amplifier has a nominal sensitivity of 25 mV/fC with a wire effective capacitance of 110 pF. The ADC is the CAEN V1724 8-channel 100 Ms/s VME unit, read out via optical CAEN\footnote{CAEN S.p.A. Via Vetraia, 11, 55049 - Viareggio (LU) - ITALY} V2718 VME-PCI interface. The typical response of the amplifier to the short ($<1\mu s$) charge packet is shown in Figure \ref{TypSignal}.

 
The triggering system is based on majority logics allowing to trigger on each wire,
as well as on the OR of any number of wires out of 8. The analog pulse is amplitude-discriminated. The collected charge descrimination threshold can be varied from 0 to 30~fC. After each trigger the data from each wire (4096 samples) are saved on disk together with meta-information, like event time, run parameters, etc. These files are then converted into a ROOT data structure and analised in the ROOT physics analysis framework\footnote{ROOT Physics Analysis Framework. http://root.cern.ch}.

\begin{figure}[htbp]
  \centering
  \begin{minipage}[b]{9.0 cm}
    \includegraphics[width=1\textwidth]{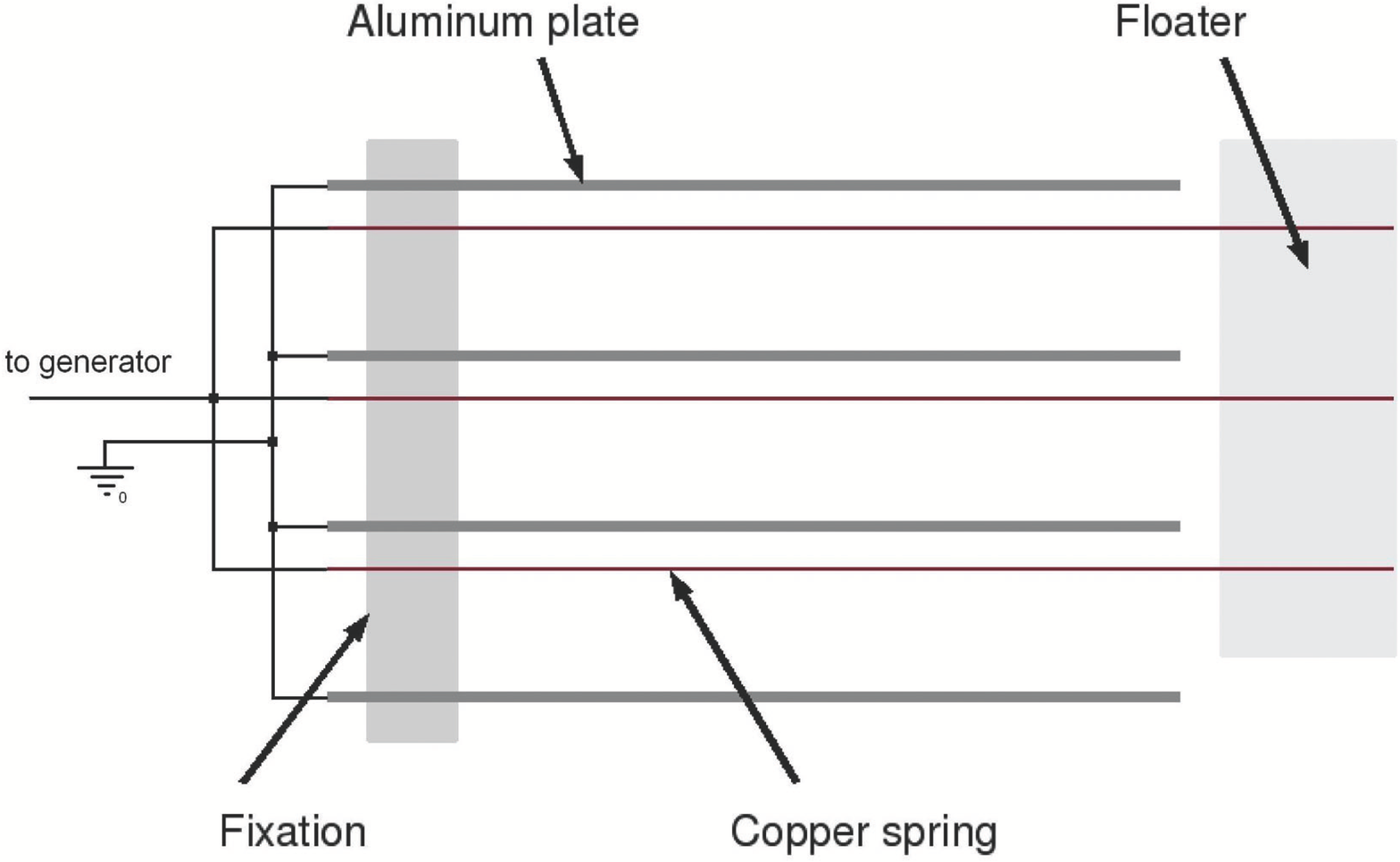}  
		\caption{Scheme of the liquid Argon-Nitrogen density meter.}
  \label{shemadensity}
  \end{minipage}
  
\end{figure}

In order to measure relatively high concentrations of Nitrogen in Argon we used a capacitive density meter. Due to the fact that Argon ($\rho = 1,38 kg/l$) is much denser than Nitrogen ($\rho = 0.809 kg/l$) the density of the mixture provides the concentration of Nitrogen in Argon. The dependence of the density on the concentration of Nitrogen in Argon is linear. In Figure~\ref{shemadensity} the structure of the density meter is shown. The density meter sensor is a capacitor with one plate attached to a floater. Being submersed into the liquid, the capacitance value depends on the buoyancy force acting on the floater and deforming the plate. In order to measure this capacitance, a relaxation generator was built with the sensor in a time-defining circuit. The frequency of the generator is then measured. The density meter was calibrated with different Argon-Nitrogen mixtures and has shown a linear dependence in the range of 0$\%$ to 100$\%$ of Nitrogen in Argon. The density measurement error is found to be about 1$\%$.   


\begin{figure}[htbp]	
\center\includegraphics[width=0.7\textwidth]{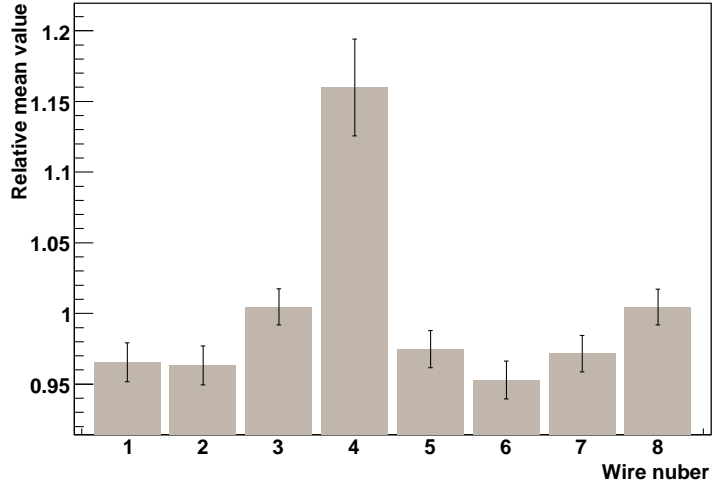}
\caption{Relative mean value of the wire signals from the $^{60}Co$ source. The TPC is filled with the pure liquid Argon.}\label{meanwire}
\end {figure}

A dedicated charge-integrating shaping preamplifier was used for measurements. Assuming that the charge packet extension is short compared to the shaper relaxation time ($1 \mu s$), the pulse height is proportional to the total collected charge. In order to normalize gain factors for different channels, the mean signal amplitude for each of them was measured by using Compton electrons from the $^{60}Co$ source. The distribution of the relative average signals for 8 wires is shown in Figure \ref{meanwire}.

\section{Experimental procedure and data taking}
According to \cite{Ikarus2} this initial Argon purity should provide a charge life time of about 0.15 $\mu s$, which for drift fields higher than 2~kV/cm results in a charge attenuation lenght of about 1 mm, comparable to the drift lenght of the TPC. 

\begin{figure}[htbp]
  \centering
  \begin{minipage}[b]{7.5 cm}
    \includegraphics[width=1\textwidth]{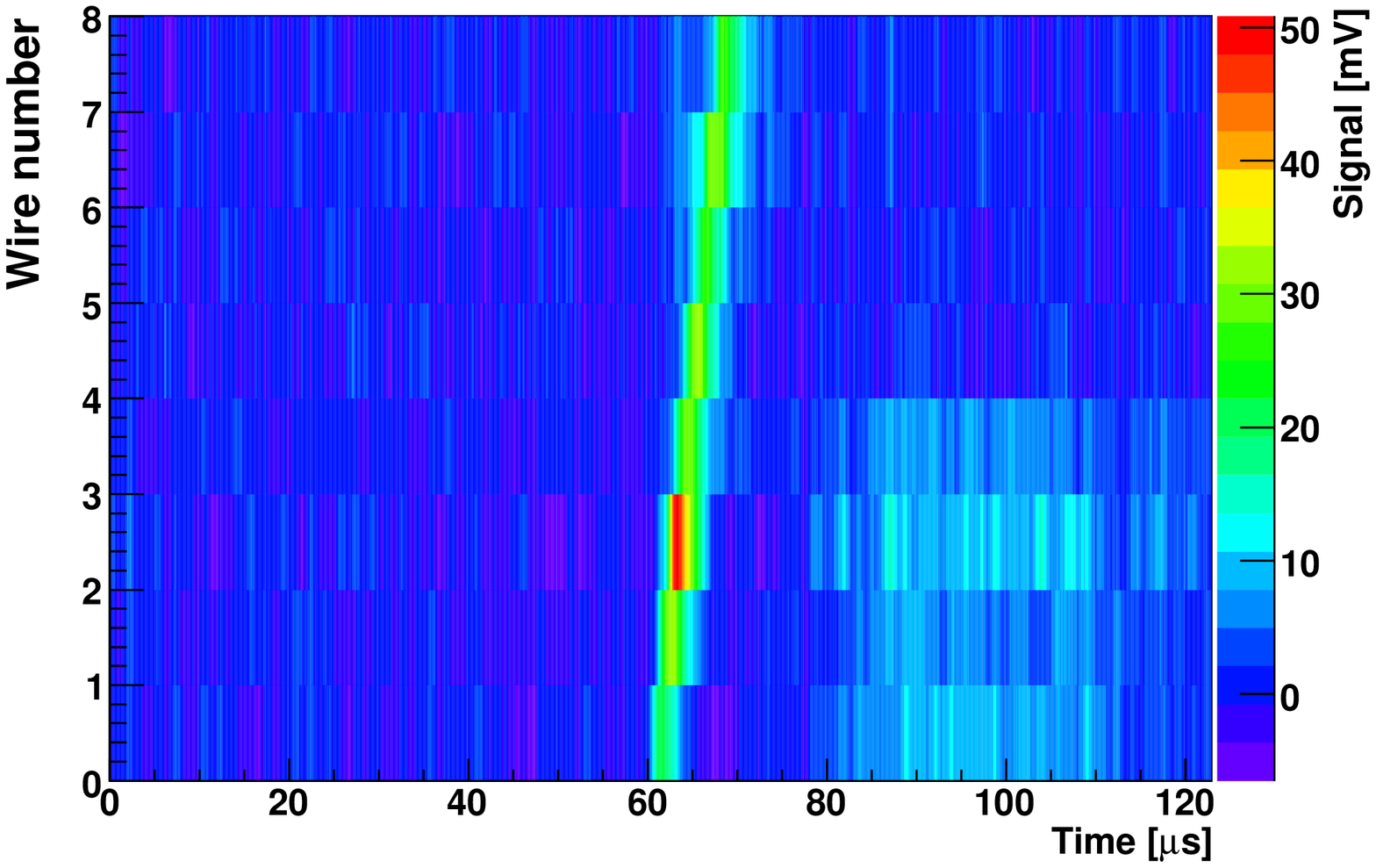}  
  \end{minipage}
  \begin{minipage}[b]{7.5 cm}
    \includegraphics[width=1\textwidth]{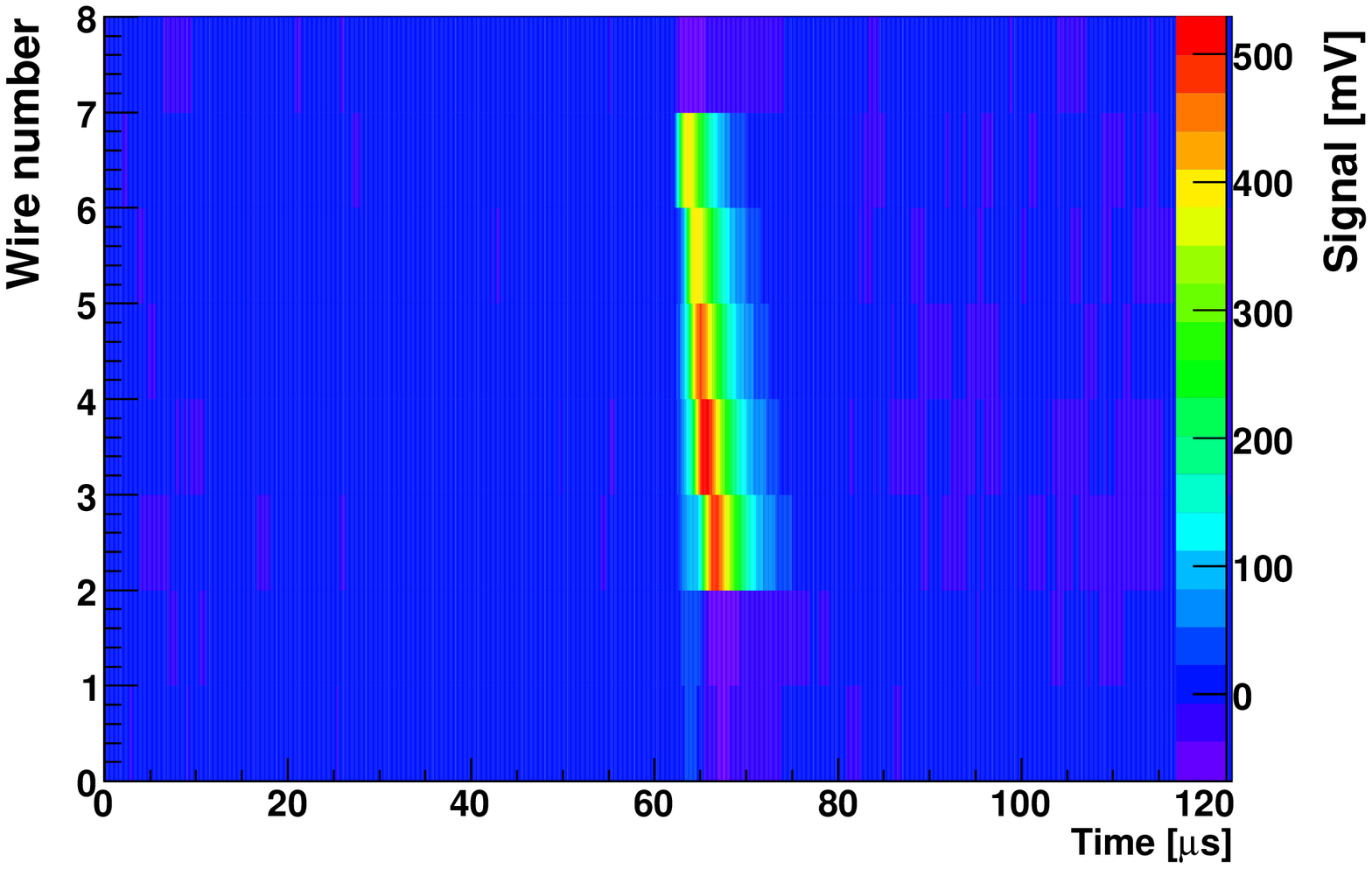}  
  \end{minipage}

  \centering
  \begin{minipage}[b]{7 cm}
    \includegraphics[width=1\textwidth]{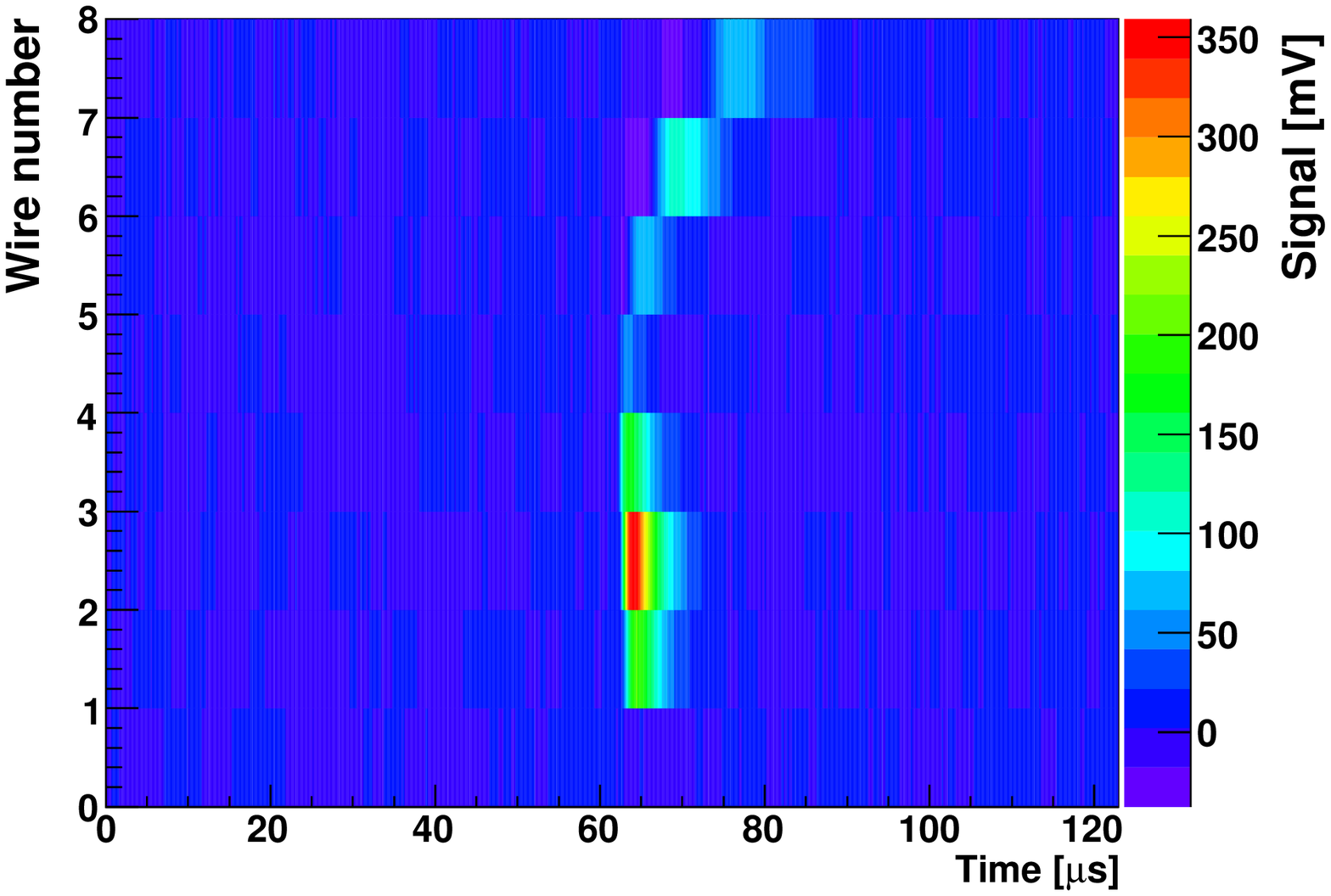}  
  \end{minipage}
  \begin{minipage}[b]{7.5 cm}
    \includegraphics[width=1\textwidth]{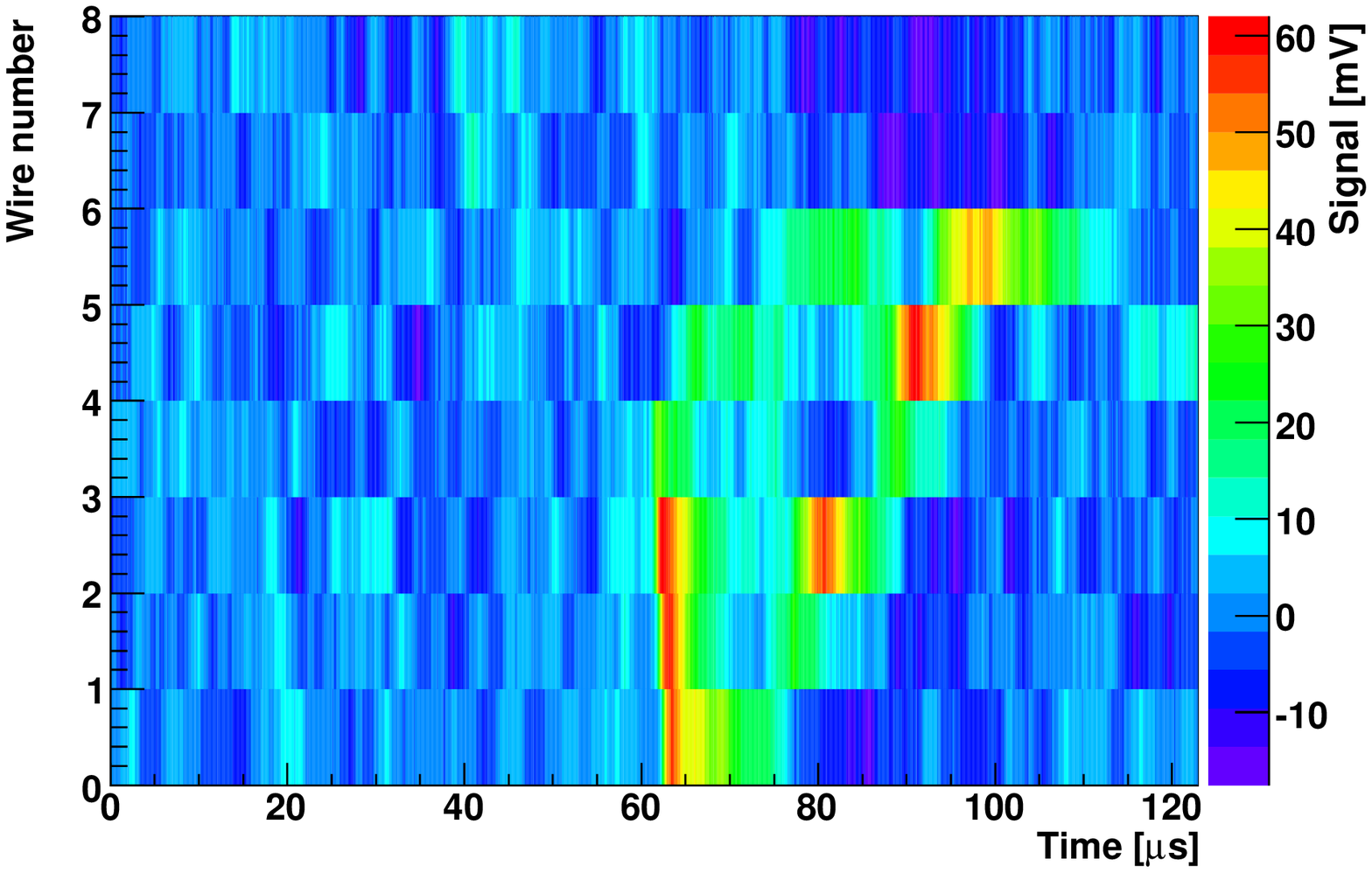}  
  \end{minipage}
  \caption{Cosmic-ray muon tracks detected by the TPC filled with pure liquid Argon.}
  \label{Cosmic2}
\end{figure}

The initial tests were done with the non-doped Argon, triggering on the coincidence of 4 wire signals. The discrimination level was set to about half of the expected pulse width from minimum ionizing particles passing perpendicular to the wires (about 3~fC). With these conditions we observed tracks from cosmic-ray muons crossing the active volume of the chamber (Figure \ref{Cosmic2}).

\begin{figure}[htbp]	
\center\includegraphics[width=0.7\textwidth]{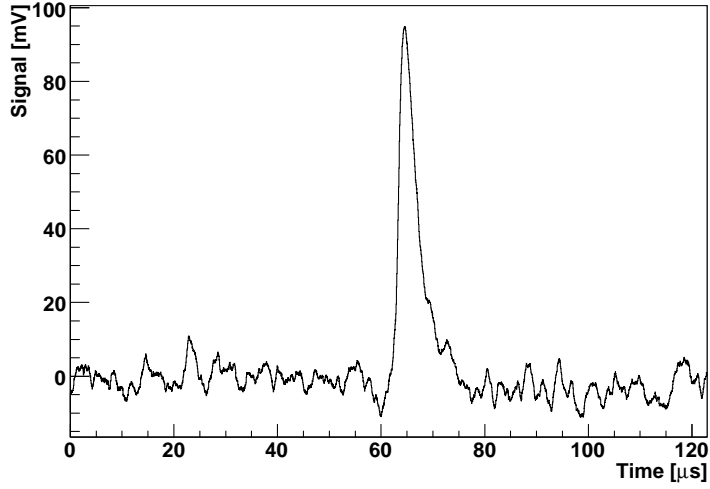}
\caption{Typical signal from $^{60}Co$ Compton ionization on one TPC wire. TPC is filled with the pure liquid Argon.}\label{TypSignal}
\end {figure}

The total collected charge represents the convolution of the produced ionization charge reduced by the initial recombination and by charge losses due to attachments on impurities during the drift. The energy stectrum of cosmic muons is rather broad and the rate of events is relatively low. Therefore, these tracks could not be efficiently used to study charge recombination in our setup. In order to get reasonable statistics a radioactive $^{60}Co$ source  was used. This source emits $\gamma$-rays in two narrow spectral lines of 1.17~MeV and 1.33~MeV with equal intensities. These $\gamma$ may experience Compton scattering on the electrons in the liquid Argon, so producing electrons with a typical Compton energy spectrum. Such electrons have a range in Argon of the order of 1mm, so the ionization charge of each of them can be only seen by one wire of the chamber. The primary ionization charge can then be calculated using the known value of the ion-electron pair production energy in liquid Argon $W=23.6~eV$ \cite{Ikarus,ThomasImel}. For the end point Compton electron from 1.33 MeV $\gamma$ the total deposited charge is expected to be $Q_0=7.56~fC$.  The typical resulting signal from the TPC wire is shown in Figure \ref{TypSignal}.

To trigger the aquisition for further measurements we used a discrimination threshold corresponding to about 1/5 of the response to the  mostly energetic Compton electrons from $^{60}Co$. In the next Section we will show how this signal can be used to decouple different charge loss mechanisms in the liquid Argon-Nitrogen mixture.

\section{Charge recombination and observed signal in the TPC}


The initial spectrum of ionization in a drift chamber is transformed by the following processes:
\begin{itemize}
\item{Recombination}


Let's assume that the probability density for the initial ionization from a single event is approximated  by a polynomial expansion on the charge range from 0 to $q_e$:
\begin{align}
P_{q_0}(x)=  &\sum_{i=0}^n A_ix^i;~~~~&0<=x<=q_e;\label{Pq0}
 \\ &0; &x>q_e~or~x<0; 
\end{align}
where $x$ represents the ionization charge and $A_i$ are expansion coefficients.

Recombination happens just after the formation of the charge packet around the particle track. There are models explaining this process \cite{Onsager,Jaffe,Kramers,ThomasImel} which however have difficulties in describing the recombination in all possible ranges of drift fields and ionization densities. The so-called "Box model" \cite{ThomasImel} gives a good theoretical basis and allows to take into account possible effects of substantial Nitrogen concentrations. The model is described by the formula: 
\begin{equation}
q_0=Q_0\frac{1}{\xi} ln(1+\xi),~~~~\xi=\frac{N_0 K_r}{4a^2u_-E},
\end{equation}
where $Q_0$ is the initial ionization charge, $q_0$ is the collected charge assuming no drift,
$N_0$ is the total amount of each (+ and -) charge in the ionization "box" of linear dimension $a$ in the direction of the electric field, $K_r$ is the recombination term defined by the recombination cross-section, $u_-$ is the electron mobility and $E$ is the electric field intensity.

\item{Charge attachment to impurities during the drift}

The charge recorded at the TPC collection wire is defined by:
\begin{equation}
q=q_0e^{-x/\lambda}
\end{equation}
where $q_0$ is the charge left after recombination at the distance $x$ from the wire
and $\lambda$, the attenuation length, is the function of the Argon purity and temperature.

\end{itemize}

The probability distribution of the collected charge in case of primary ionization by Compton electrons is given by:

\begin{enumerate}
\item{A. $0<x<q_e\,e^{-D/\lambda}$ }
\begin{equation}
P_q(x)=\frac{\lambda}{D} \left[  \frac{8}{3} (e^{D/\lambda}-1) - \frac{8}{3} (e^{2D/\lambda}-1) \frac{x}{E_\gamma} + \frac{14}{9} (e^{3D/\lambda}-1) \frac{x^2}{E_\gamma^2} +\frac{1}{2}  (e^{4D/\lambda}-1)\frac{x^3}{E_\gamma^3} \right];
\label{CsA}
\end{equation}
\item{B. $q_e\,e^{-D/\lambda}<x<q_e$}
\begin{equation}
P_q(x)=\frac{\lambda}{D} \left[  \frac{8}{3} (\frac{q_e}{x}-1)- \frac{8}{3 E_\gamma} (\frac{q_e}{x}q_e-x)+  \frac{14}{9 E_\gamma^2} (\frac{q_e}{x}q_e^2-x^2) +\frac{1}{2 E_\gamma^3} (\frac{q_e}{x}q_e^3-x^3)  \right];
\label{CsB}
\end{equation}
\end{enumerate}

where $D$ is the distance between the anode wires and the cathode plate, $\lambda$ is the charge attenuation length during the drift, $E_\gamma$ is the energy of the primary $\gamma$ and $q_e$ is the end point of the Compton electron energy spectrum.

\begin{figure}[htbp]
\center\includegraphics[width=.5\textwidth]{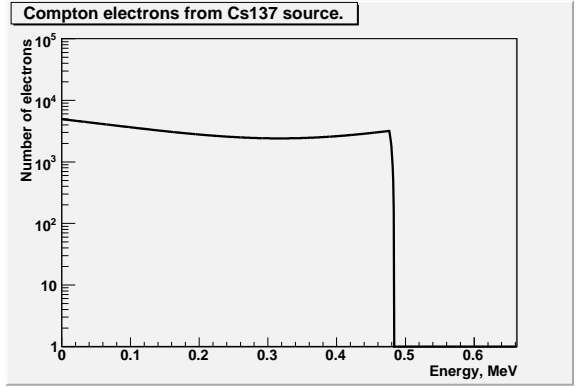}
\caption{Non-normalized calculated distribution for the energy of Compton-scattered electrons from $^{137}Cs$ $\gamma$ in liquid Argon.}
\label{fig1}
\end{figure}

\begin{figure}[htbp]
\center\includegraphics[width=.5\textwidth]{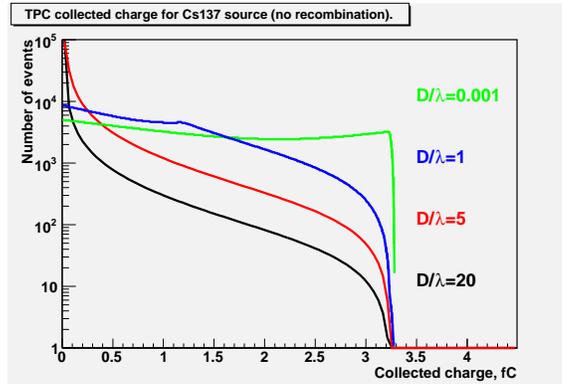}
\caption{Calculated collected charge for Compton-scattered electrons from $^{137}Cs$ $\gamma$ in liquid Argon.}
\label{fig2}
\end{figure}

From Figure \ref{fig2} it is evident that the end-point of the spectrum does not depend on the drift attenuation. Therefore, formulas \ref{CsA} and \ref{CsB} can be used to fit the TPC signal spectrum in order to derive the charge left after the initial recombination $q_0$, and so to decouple it from the subsequent drift losses. In this work a $^{60}Co$ isotope with two lines in the $\gamma$ spectrum is used and the fitting function is modified accordingly.

\section{Recombination in liquid Argon doped with Nitrogen}

\begin{figure}[htbp]	
\center\includegraphics[width=1\textwidth]{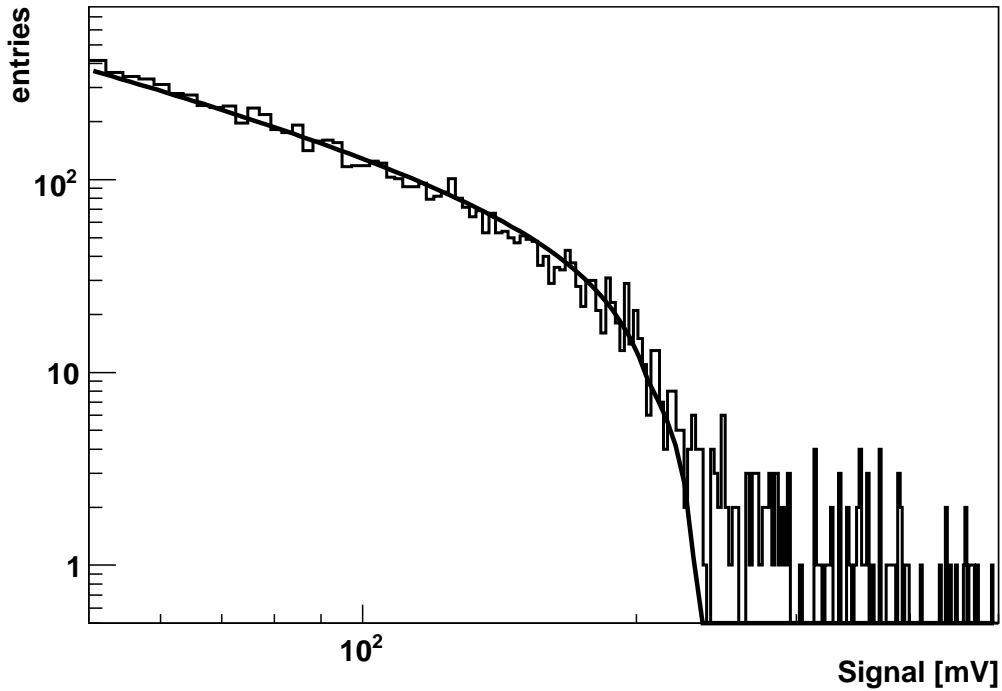}
\caption{Compton spectrum of the $^{60}Co$ at 2.94 kV/cm drift field intensity. The fitting curve is given by formulas 4.5 and 4.6 in the text. TPC is filled with the pure liquid Argon.}\label{Co60spect}
\end {figure}

\begin{figure}[htbp]	
\center\includegraphics[width=.9\textwidth]{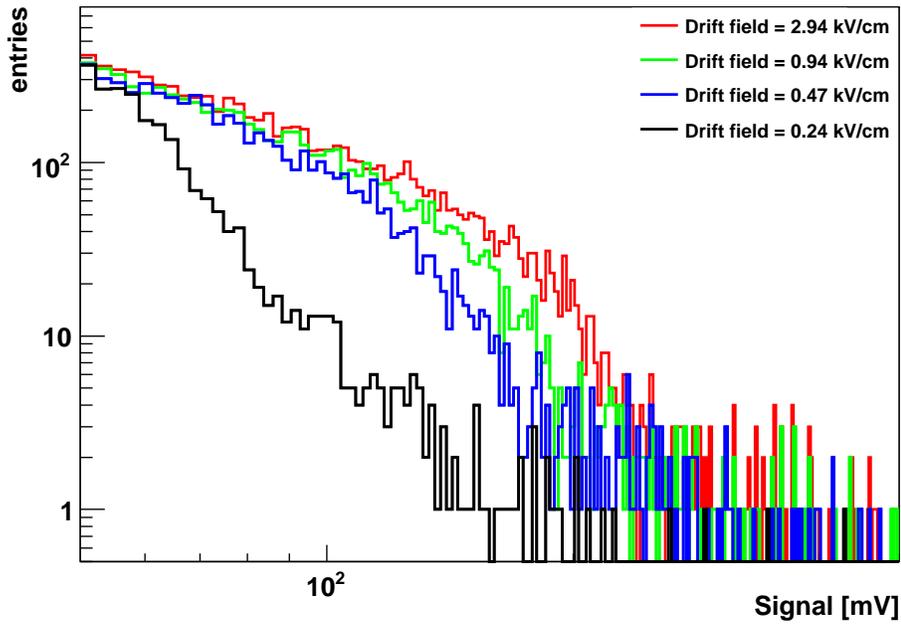}
\caption{Compton spectra of the $^{60}Co$ in pure Argon for different drift field intensities.}\label{Co60diffE}
\end {figure}

\begin{figure}[htbp]	
\center\includegraphics[width=.9\textwidth]{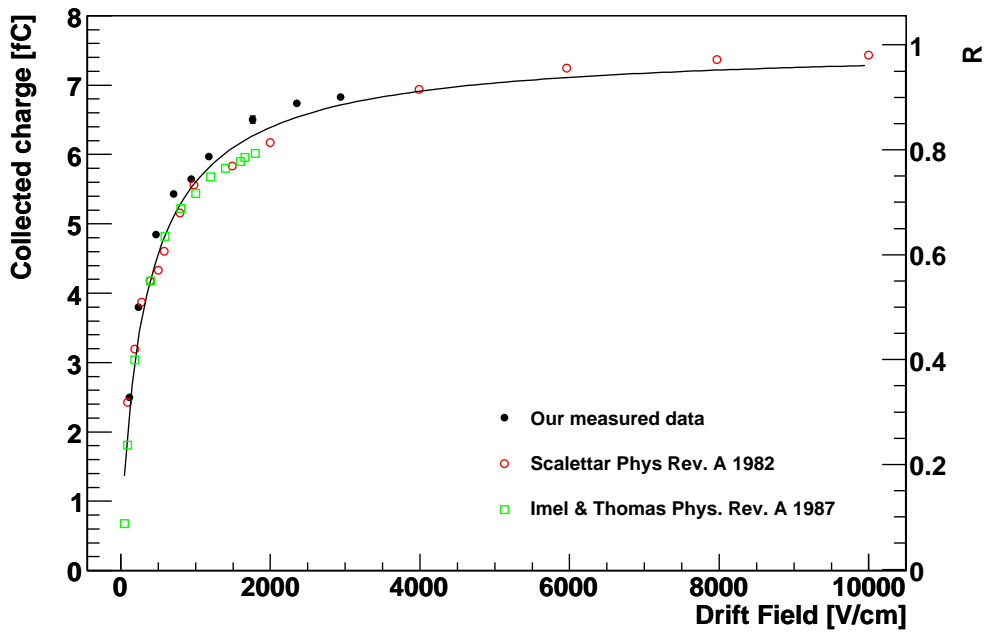}
\caption{Compton edge of the $^{60}Co$ in pure Argon compared with other experiments. The curve represents a fit with the "Box model".}\label{Co60diffexperiments}
\end {figure}

In order to extract information about the charge left after recombination, the pulse height distribution from the $^{60}Co$ source was fitted with the superposition of functions \ref{CsA} and \ref{CsB} for two values of $q_e$, corresponding to $\gamma$ energies of 1.17~MeV and 1.33~MeV (Figure \ref{Co60spect}). Measurements with non-doped Argon were made first in order to obtain reference data. Figure \ref{Co60diffE} shows the signal spectra for four different values of the drift field in the TPC. After fitting, the end-point value can be plotted versus the drift field intensity. Figure \ref{Co60diffexperiments} shows the end point values, superimposed to the "Box model" curve. The $Q_0$ value derived from the fit corresponds to the maximum charge that can be produced by $^{60}Co$ Compton electrons: $Q_0=7.56 fC$, as mentioned above. This value is used to calibrate the sensitivity of the charge amplifiers. In Figure \ref{Co60diffexperiments} data from \cite{ThomasImel} are also shown, scaled by the ratio between the energy of the $\beta$-source electrons used in that work ($364~keV~e^-~$from$~^{113}Sn$) and the maximum energy of Compton electrons in our conditions ($1.1~MeV$), showing good agreement.

\begin{figure}[htbp]	
\center\includegraphics[width=.8\textwidth]{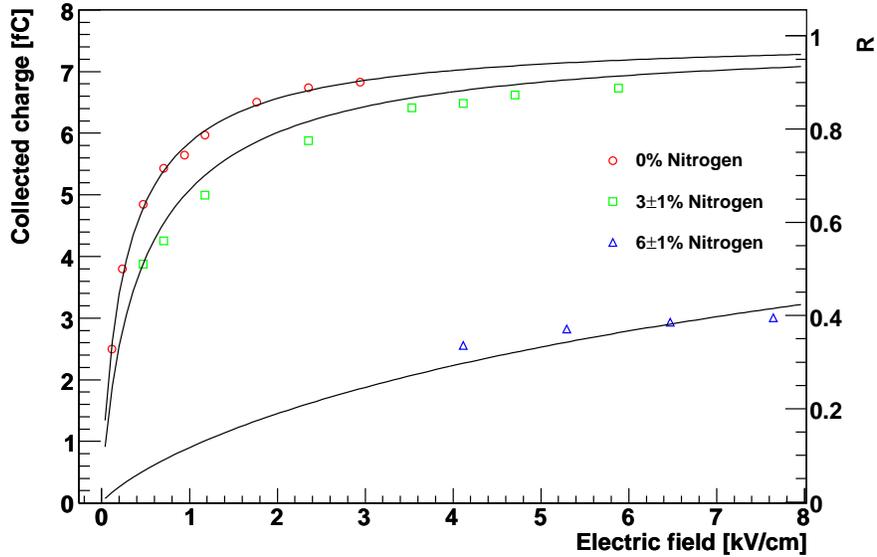}
\caption{The charge corresponding to the Compton edge (collected charge assuming no drift $q_0$) from $^{60}Co$ for different drift field intensities. Points are fitted with the "Box model", assuming the initial ionization charge $Q_0$ constant and equal to the one for non-doped Argon. On the right vertical axis the charge value relative to the fitted $Q_0$ is shown: $R=\frac{q_0}{Q_0}$}\label{c2}
\end {figure}

\begin{figure}[htbp]	
\center\includegraphics[width=.8\textwidth]{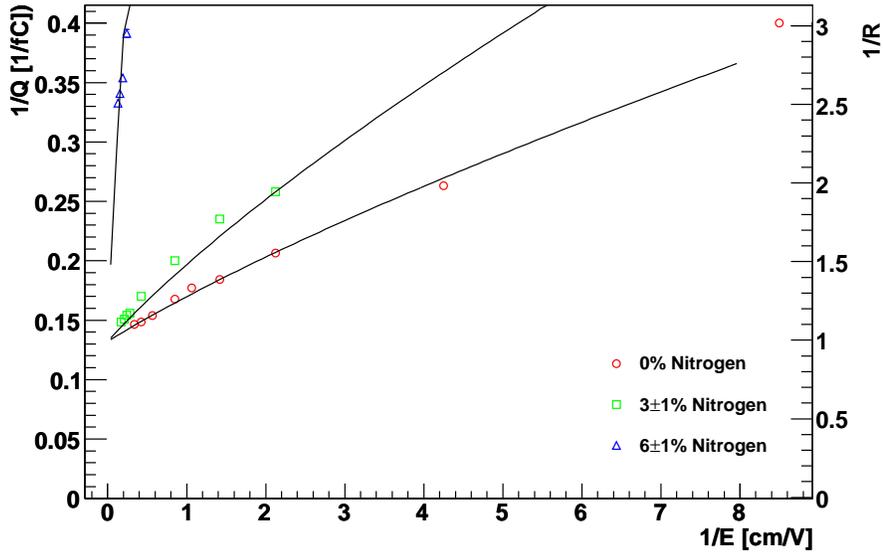}
\caption{The charge corresponding to the Compton edge (collected charge assuming no drift $q_0$) from $^{60}Co$ for different drift field intensities. Points are fitted with the "Box model", assuming the initial ionization charge $Q_0$ constant and equal to the one for non-doped Argon. The data are represented in reversed $\frac{1}{R}$ vs $1/E$ coordinates.}\label{c3}
\end {figure}

\begin{figure}[htbp]	
\center\includegraphics[width=.8\textwidth]{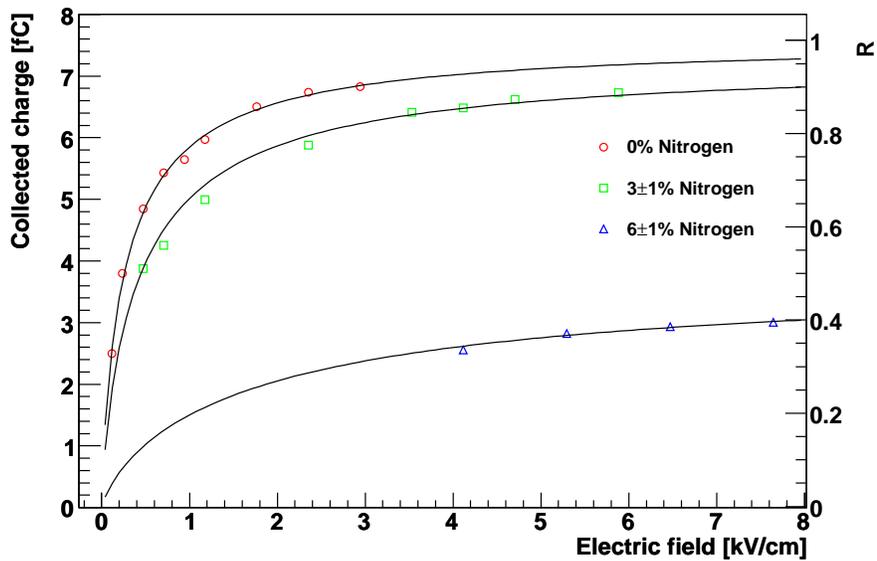}
\caption{Data fitted with additional charge quenching factor $\alpha$.}\label{c4}
\end {figure}

\begin{figure}[htbp]	
\center\includegraphics[width=.8\textwidth]{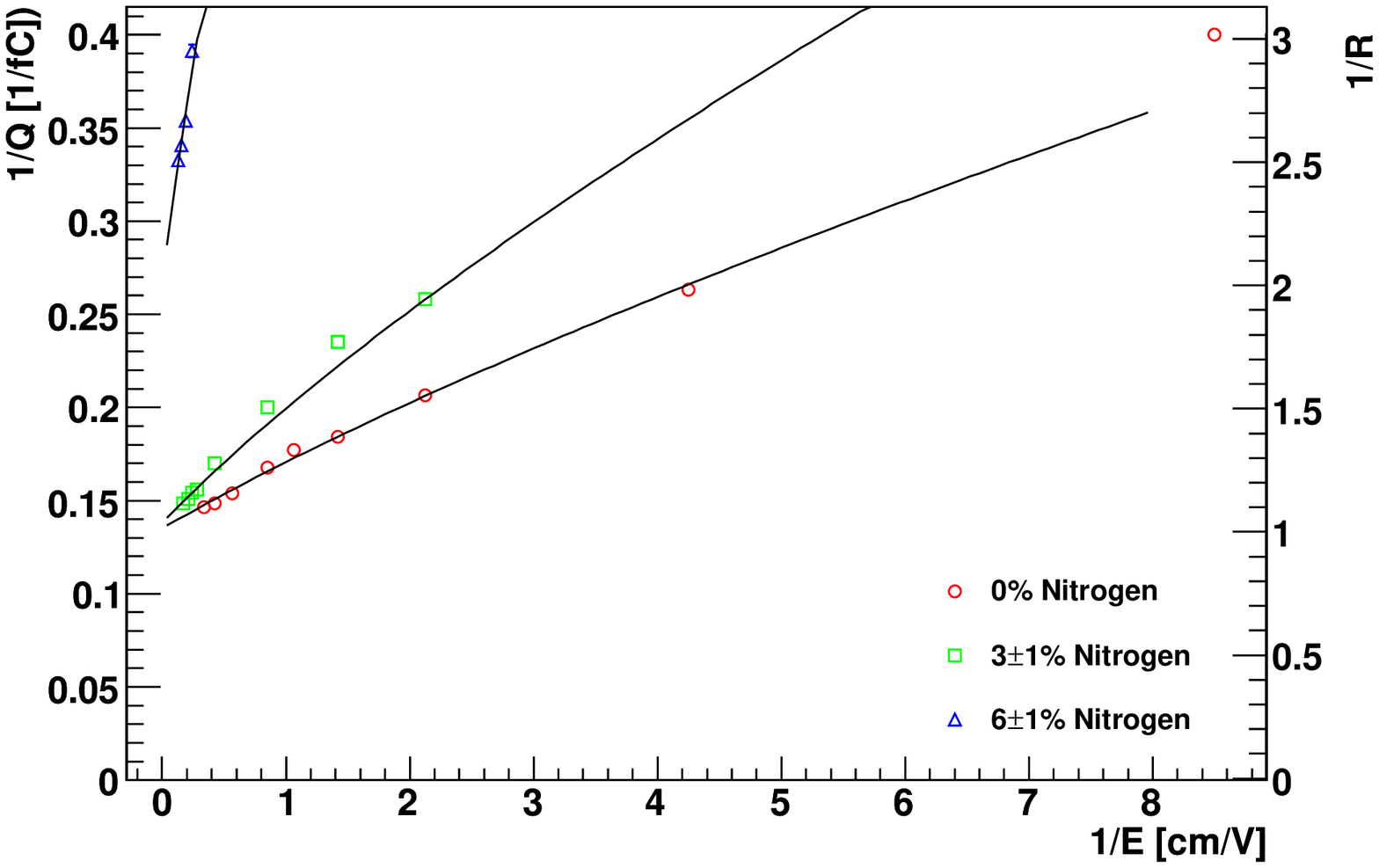}
\caption{Data fitted with an additional charge quenching factor $\alpha$ represented in reversed $\frac{1}{R}$ vs $1/E$ coordinates.}\label{c5}
\end {figure}

Liquid Nitrogen was then gradually added to the cryostat containing liquid Argon. Its content was monitored by the density meter. Two sets of measurements were done for Nitrogen concentrations of $3\pm1 \%$ and $6\pm 1\%$ respectively. The dependence of $q_0$ on the drift field is shown in Figure \ref{c2}. In Figure \ref{c3} the same data are plotted in 
reversed coordinates: $\frac{1}{q_0/Q_0}$ vs $1/E$, in order to demostrate the nearly linear behaviour.
Data points are fitted with the "Box model":
\begin{equation}
q_0=Q_0\frac{1}{\xi} ln(1+\xi),~~~~\xi=\frac{\beta}{E},
\end{equation}
where $Q_0$ is the initial ionization charge for pure Argon, $q_0$ is the collected charge assuming no drift,
$\beta$ is the fitting parameter and $E$ is the electric field intensity. From these Figures it is evident that the model with a constant $Q_0$ measured for pure Argon does not fit well the experimental data with doped Argon. Introducing an additional factor $\alpha$ in front of $Q_0$ helps in matching data much better (see Figures \ref{c4} and \ref{c5}): 
\begin{equation}
q_0=\alpha \times Q_0\frac{1}{\xi} ln(1+\xi),~~~~\xi=\frac{\beta}{E},
\end{equation}

\begin{figure}[htbp]
  \centering
  \begin{minipage}[b]{10.3 cm}
    \includegraphics[width=.9\textwidth]{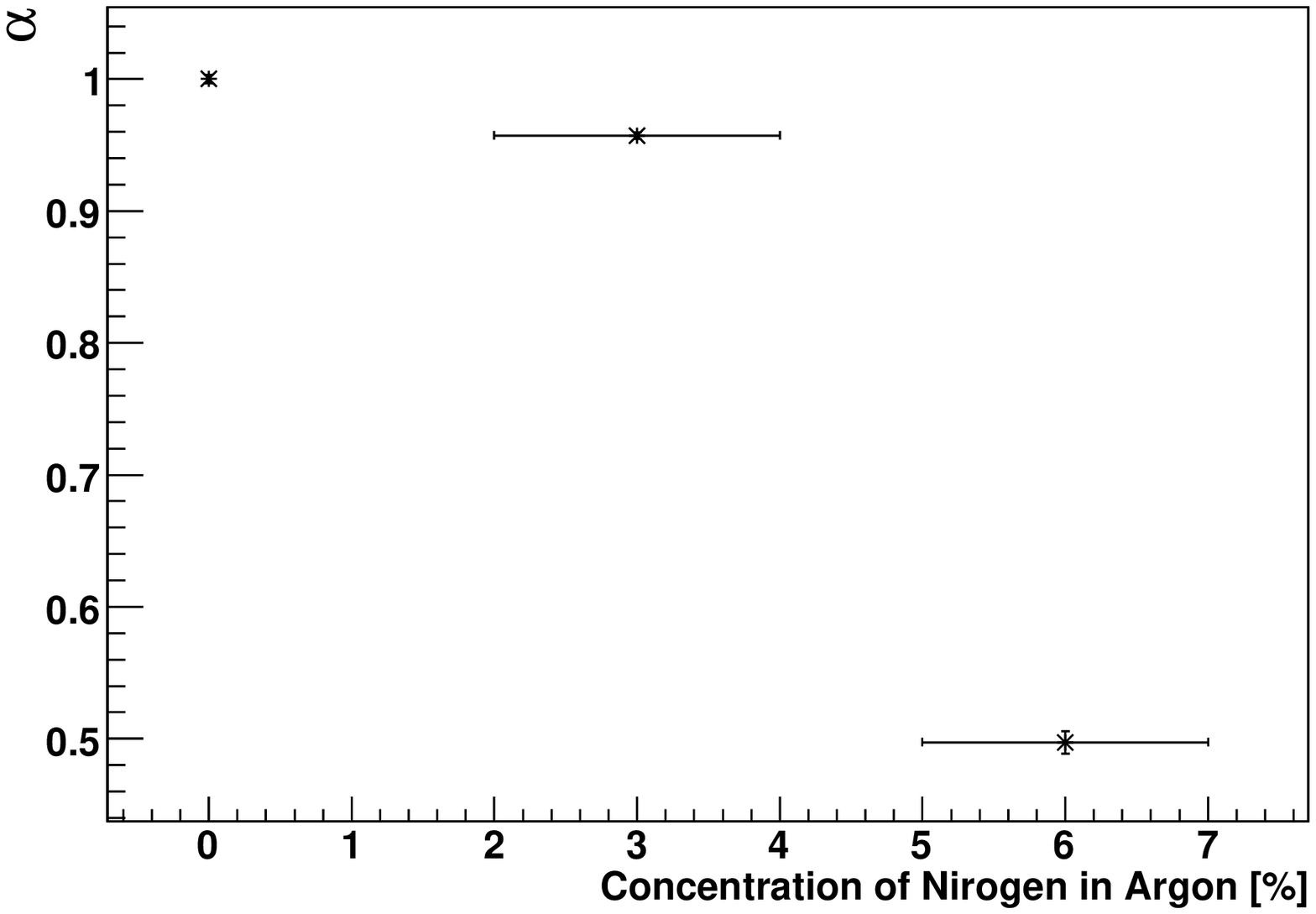}  
		\caption{$\alpha$ parameter as a function of Nitrogen concentration.}
  \label{Alpha}
  \end{minipage}
  \begin{minipage}[b]{10.3 cm}
    \includegraphics[width=.9\textwidth]{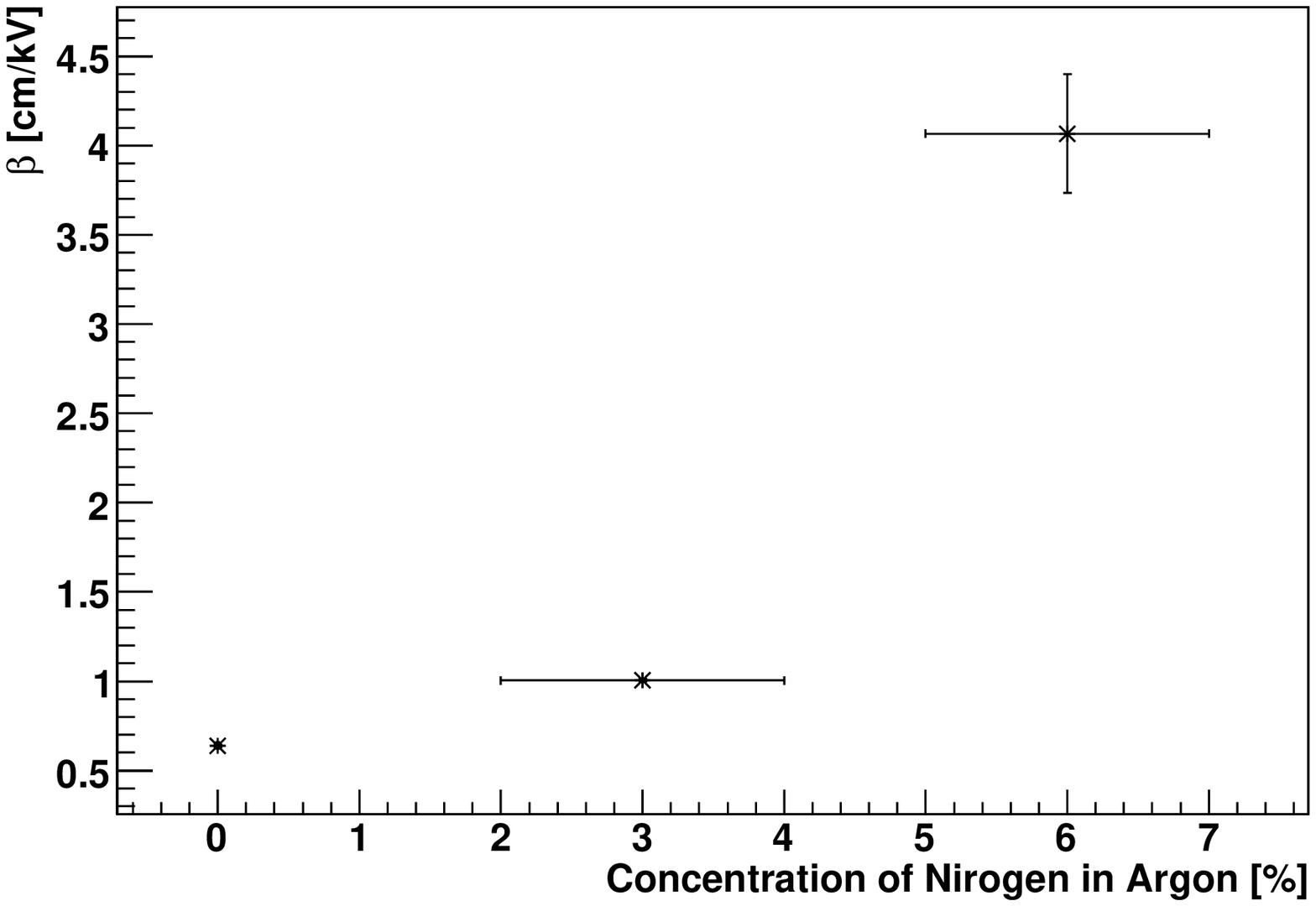}  
  
  \caption{$\beta$ parameter as a function of Nitrogen concentration.}
  \label{Beta}
	\end{minipage}
\end{figure}

Figures \ref{Alpha} and \ref{Beta} show the dependence of these fitted parameters versus the Nitrogen content. The factor $\alpha$ being less than 1 for Argon doped with Nitrogen can be due to the fact that the energy needed to create an ion-electron pair in the mixture is higher than for pure Argon ($W=23.6~eV$). The ionization energies of Argon and Nitrogen are very close to each other. The strong observed effect on $\alpha$ can be explained by the "quenching" of the ionization during production, that is even before the primary recombination phase. The presence of rotational and vibrational energy levels of molecular Nitrogen of the order of 2-10~eV may enhance electron thermalization during the ionization avalanche, therefore decreasing the total charge yield \cite{Takahashi}. In fact, this effect was already observed by D.W. Swan \cite{Swan} for lower concentrations of Nitrogen (below $1\%$), but no reasonable theoretical explanation was given. This effect is field-independent, so the charge loss can not be compensated by the drift field.

The $\beta$ value in the "Box model" is derived as: 
\begin{equation}
\beta=\frac{N_0~k_{r}}{4a^2u_-},
\end{equation}
where $N_0$ is the number of produced electron-ion pairs in the volume of linear dimension $a$, $k_{r}$ is referred to as a recombination rate term, and $u_-$ is the electron mobility. The recombination rate term is in particular proportional to the collision cross-section between the corresponding ions and the electrons.
Significant growth of the $\beta$ value can thus be attributed to the increase of $k_{r}$ in the mixture, compared to pure Argon, or to a decrease of the electron mobility $u_-$ in the mixture. However, the latter was reported to show opposite behaviour, at least for small Nitrogen concentrations (below $1\%$)\cite{Swan}.

Detailed theoretical calculations were recently made for dense gaseous mixtures of Krypton and Nitrogen \cite{Wojcik}. These calculations show a strong dependence of the recombination rate term on the Nitrogen concentration up to $5~mol.\%$, very similar to the one we observed. The effect is explained by a much lower mean kinetic energy of the electron in presence of vibrational and rotational molecular Nitrogen energy levels. The recombination happens only if the total relative energy of the electron-ion pair falls below a certain threshold. Therefore, decreasing the mean energy enhances the recombination rate. This mechanism was suggested in \cite{Barabash}, where the capture probability of electrons on electronegative impurities in liquid Argon-Nitrogen mixtures was studied. In liquid Argon strong dependence of the electron capture probability on the Nitrogen concentration was shown. The electron capture on impurity atoms has a similar nature as the recombination. Moreover, the dependance of the capture probability on the Nitrogen concentration that was observed in \cite{Barabash} agrees with the asymptotic behaviour of recombination rate parameter $\beta$ obtained by us (Figure \ref{Beta}). However, as was mentioned in \cite{Barabash}, this effect can be compensated by increasing the drift field intensity.

\section{Conclusions}

Ionization signals from cosmic muons and Compton electrons have been observed in a liquid Argon TPC, doped with substantial concentrations of Nitrogen. The so-called "Box model" describing electron-ion recombination can be used to fit our data assuming that the recombination parameter $\beta$ depends on the Nitrogen concentration, and that some additional field-independent charge quenching is present. Our results prove the feasibility of liquid Argon, Nitrogen doped TPCs for applications that require a Nitrogen content in the tracking medium, such as GRNA radiography. The charge response in such mixtures will be further studied in following publications.

\section*{Acknowledgements}
We wish to warmly thank all our technical collaborators for their skillful help in building and operating the detector and the related infrastructure.


\begin{thebibliography}{99}

\bibitem{Rubbia77} C. Rubbia, "The Liquid Argon Time Projection Chamber: a New Concept for Neutrino Detector", CERN-EP 77-08 (1977).

\bibitem{Ikarus1} F. Arneodo et al.,"Observation of long ionizing tracks with the ICARUS T600 first half-module" Nucl. Inst. Meth., A508 (2003) 287-294.

\bibitem{GRA1} D. Vartsky et al., "Gamma Ray Nuclear Resonance Absorption: An Alternative Method for in Vivo Body Composition Studies",  Annals of the New York Academy of Sciences 904:236-246 (2000).

\bibitem{GRA2} G. Feldman et al., "Analysis of gamma-ray nuclear resonant absorption (NRA) images for automatic explosives detection", Image Processing and Its Applications, 1999. Seventh International Conference on (Conf. Publ. No. 465)
Volume 2, Issue , 1999 Page(s):789 - 793 vol.2.

\bibitem{DavidsonLarsh} N. Davidson and A. E. Larsh, Jr.,"Conductivity Pulses in Liquid Argon", Phys. Rev. 74, 220 - 220 (1948).

\bibitem{Swan} D. W. Swan, "Electron Attachment Processes in Liquid Argon containing Oxygen or Nitrogen Impurity",  1963 Proc. Phys. Soc. 82 74-84.

\bibitem{Hofmann} W. Hofmann et al.? "Production and transport of conduction electrons in a liquid Argon ionisation chamber", Nucl. Inst. Meth., A135 (1976) 151-156

\bibitem{Berset} J. C. Berset et al., "Scintillation light from liquid Argon and its use in a new hybrid detector", Nucl. Inst. Meth., A203 (1982) 133-140 

\bibitem{Barabash} A.S. Barabash et al., "Investigation of electronic conductivity of liquid Argon-Nitrogen mixtures",  Nucl. Inst. Meth., A234 (1985) 451-454

\bibitem{Sakai}  T. Kimura, Y. Sakai, H. Tagashira, S. Nakamura, "Fast and slow electrons in liquid Ar and N2 mixtures", IEEE Transactions on Dielectrics and Electrical Insulation, 
Vol. 1, Issue 4, Aug 1994 Page(s):644 - 647


\bibitem{Onsager} L. Onsager, "Initial Recombination of Ions", Phys. Rev. 54, 554 - 557 (1938).

\bibitem{Jaffe} G. Jaffe, "Zur theorie der ionization in kolonnen", Ann. Phys. 42303-344 (1913).

\bibitem{Kramers} H.A. Kramers, "On a modification of Jaffe's theory of column-ionization", Physica
Volume 18, Issue 10, October 1952, Pages 665-675.  

\bibitem{ThomasImel} J. Thomas and D. A. Imel, "Recombination of electron-ion pairs in liquid argon and liquid xenon", Phys. Rev. A 36, 614 - 616 (1987).

\bibitem{Ikarus} S. Amoruso et. al., "Study of electron recombination in liquid Argon with the ICARUS TPC", Nucl. Instr. and Meth. A523, Issue 3, 11 May 2004, 275-286  

\bibitem{Ikarus2} E. Buckley et al., "A study of ionization electrons drifting large distances in liquid argon", Nucl. Instr. and Meth. A275, (1989), 364.

\bibitem{Takahashi} T. Takahashi et al, "Electron thermalization in Argon-Nitrogen gas mixture excited by $^{252}Cf$ fission fragments", Phys. Rev. A, Vol.25, Number 5, pp.2820-2823, May 1982. 

\bibitem{Wojcik} Wojcik M., Tachiya M., "Effect of molecular additives on electron mobility and electron-ion recombination rate constant in dense gaseous krypton", Journal of physical chemistry A, 2002, vol. 106, no18, pp. 4468-4475.


\end{thebibliography}
\end{document}